\documentclass[aps,prd,twocolumn,superscriptaddress,preprintnumbers,nofootinbib]{revtex4-2}
\usepackage{scalerel}
\usepackage{tikz}
\usetikzlibrary{svg.path}

\definecolor{orcidlogocol}{HTML}{A6CE39}
\tikzset{
  orcidlogo/.pic={
    \fill[orcidlogocol] svg{M256,128c0,70.7-57.3,128-128,128C57.3,256,0,198.7,0,128C0,57.3,57.3,0,128,0C198.7,0,256,57.3,256,128z};
    \fill[white] svg{M86.3,186.2H70.9V79.1h15.4v48.4V186.2z}
                 svg{M108.9,79.1h41.6c39.6,0,57,28.3,57,53.6c0,27.5-21.5,53.6-56.8,53.6h-41.8V79.1z M124.3,172.4h24.5c34.9,0,42.9-26.5,42.9-39.7c0-21.5-13.7-39.7-43.7-39.7h-23.7V172.4z}
                 svg{M88.7,56.8c0,5.5-4.5,10.1-10.1,10.1c-5.6,0-10.1-4.6-10.1-10.1c0-5.6,4.5-10.1,10.1-10.1C84.2,46.7,88.7,51.3,88.7,56.8z};
  }
}
\newcommand{\redmapper}{redMaPPer}
\newcommand\orcidicon[1]{\href{https://orcid.org/#1}{\mbox{\scalerel*{
\begin{tikzpicture}[yscale=-1,transform shape]
\pic{orcidlogo};
\end{tikzpicture}
}{|}}}}

\usepackage{natbib}
\usepackage{textcomp}
\usepackage{bm,color}
\usepackage{verbatim}
\usepackage{ulem}
\definecolor{linkcolor}{rgb}{0.6,0,0}
\definecolor{citecolor}{rgb}{0,0,0.75}
\definecolor{urlcolor}{rgb}{0.12,0.46,0.7}
\usepackage[breaklinks, colorlinks, urlcolor=urlcolor, linkcolor=linkcolor,citecolor=citecolor,pdfencoding=auto]{hyperref}
\usepackage{booktabs}
\usepackage{bigstrut}
\usepackage{xspace}
\usepackage{lineno}
\usepackage{multirow}
\usepackage{amsmath}
\usepackage{graphicx}
\usepackage{threeparttable}
\usepackage{subfigure}
\usepackage{enumitem}
\usepackage{appendix}
\usepackage{rotating}

\newcommand{\orcid}[1]{\href{https://orcid.org/#1}{\textcolor[HTML]{A6CE39}{\aiOrcid}}}

\newcommand{\nver}{\hat{\mathbf{n}}}

\newcommand{\vanish}[1]{}

\newcommand{\rver}{\hat{\mathbf{r}}}

\newcommand{\lcdm}{\ensuremath{\Lambda {\rm CDM}}\xspace}
\newcommand{\lcdmg}{\ensuremath{\Lambda {\rm CDM} + \gamma}\xspace}
\newcommand{\wcdm}{\ensuremath{w{\rm CDM}}\xspace}
\newcommand{\Planck}{\textit{Planck}\xspace}
\newcommand{\planck}{\textit{Planck}\xspace}
\newcommand{\beq}{\begin{equation}}
\newcommand{\eeq}{\end{equation}}
\newcommand{\bea}{\begin{eqnarray}}
\newcommand{\eea}{\end{eqnarray}}

\newcommand{\sr}[1]{\textcolor{red}{[SR: #1]}}

\begin{document}
\newcommand{\taupksz}{\ensuremath{(2.97 \pm 0.73) \times 10^{-3}}}
\newcommand{\cmbstn}{36}
\newcommand{\stnzbins}{(12, 20, 20, 10, 8)}
\newcommand{\tautsz}{\ensuremath{(2.51 \pm 0.55^{\text{stat}} \pm 0.15^{\rm syst}) \times 10^{-3}}}

\title{Constraining cosmological parameters using the pairwise kinematic Sunyaev-Zel’dovich effect with CMB-S4 and future galaxy cluster surveys}

\author{E.~Schiappucci \orcidicon{0000-0003-1053-8245}}
\affiliation{School of Physics, University of Melbourne, Parkville, VIC 3010, Australia}

\author{S.~Raghunathan \orcidicon{0000-0003-1405-378X}}
\affiliation{Center for AstroPhysical Surveys, National Center for Supercomputing Applications, Urbana, Illinois 61801, USA}

\author{C.~To \orcidicon{0000-0001-7836-2261}}
\affiliation{Center for Cosmology and Astro-Particle Physics, The Ohio State University, Columbus, OH 43210, USA}

\author{F.~Bianchini \orcidicon{0000-0003-4847-3483}}
\affiliation{Kavli Institute for Particle Astrophysics and Cosmology, Stanford University, 452 Lomita Mall, Stanford, CA, 94305, USA}
\affiliation{Department of Physics, Stanford University, 382 Via Pueblo Mall, Stanford, CA, 94305, USA}
\affiliation{SLAC National Accelerator Laboratory, 2575 Sand Hill Road, Menlo Park, CA, 94025, USA}

\author{C.~L.~Reichardt \orcidicon{0000-0003-2226-9169}}
\affiliation{School of Physics, University of Melbourne, Parkville, VIC 3010, Australia}

\author{N.~Battaglia \orcidicon{0000-0001-5846-0411}}
\affiliation{Department of Astronomy, Cornell University, Ithaca, NY 14853, USA}

\author{B. Hadzhiyska \orcidicon{0000-0002-2312-3121}}
\affiliation{Physics Division, Lawrence Berkeley National Laboratory, Berkeley, CA 94720, USA}
\affiliation{Berkeley Center for Cosmological Physics, Department of Physics, University of California, Berkeley, CA 94720, USA}

\author{S.~Kim \orcidicon{0000-0002-7558-8502}}
\affiliation{Department of Astronomy and Space Science, Sejong University, 209, Neungdong-ro, Gwangjin-gu, Seoul, 05006, Republic of Korea}
\affiliation{Department of Physics and Astronomy, Sejong University, 209, Neungdong-ro, Gwangjin-gu, Seoul, 05006, Republic of Korea}

\author{J.B.~Melin}
\affiliation{Université Paris-Saclay, CEA, Département de Physique des Particules, 91191, Gif-sur-Yvette, France}

\author{C.~Sif\'on \orcidicon{0000-0002-8149-1352}}
\affiliation{Instituto de Física, Pontificia Universidad Catolica de Valparaíso, Casilla 4059, Valparaíso, Chile}

\author{E.~M.~Vavagiakis \orcidicon{0000-0002-2105-7589}}
\affiliation{Department of Physics, Duke University, Durham, NC 27710, USA}
\affiliation{Department of Physics, Cornell University, Ithaca, NY 14853, USA}

\collaboration{For the CMB-S4 Collaboration}

%\author{The CMB-S4 Collaboration} 

%\linenumbers
\begin{abstract}

We present a forecast of the pairwise kinematic Sunyaev-Zel'dovich (kSZ) measurement that will be achievable with the future CMB-S4 experiment. CMB-S4 is the next stage for ground-based cosmic microwave background experiments, with a planned wide area survey that will observe approximately $50\%$ of the sky. 
We construct a simulated sample of galaxy clusters that have been optically selected in an LSST-like survey and have spectroscopic redshifts. 
For this cluster sample, we predict that CMB-S4 will reject the null hypothesis of zero pairwise kSZ signal at $\cmbstn \,\sigma$. 
We estimate the effects of systematic uncertainties such as scatter in the mass-richness scaling relation and cluster mis-centering. 
We find that these effects can reduce the signal-to-noise ratio of the CMB-S4 pairwise kSZ measurement by $20\%$.
We explore the constraining power of the measured kSZ signal in combination with measurements of the galaxy clusters' thermal SZ emission on two extensions to the standard cosmological model.
The first extension allows the dark energy equation of state $w$ to vary. 
We find the CMB-S4 pairwise kSZ measurement yields a modest reduction in the uncertainty on $w$ by a factor of 1.36 over the \Planck's 2018 uncertainty. 
%:  \lcdmg, with $\gamma$ being the growth rate which can constrain alternative theories to General Relativity, and \wcdm models. 
The second extension tests General Relativity by varying the growth index $\gamma$.
We find that CMB-S4's pairwise kSZ measurement will yield a $28\sigma$ constraint on $\gamma$, and strongly constrain alternative theories of gravity. 

%. These effects combined can dilute the signal up to 20\%, which reduces the constraining power of the pairwise kSZ by $\sim3\%$ for \lcdmg and $\sim15\%$ for \wcdm. 
\end{abstract}


\maketitle

%%%%%%%%%%%%%%%%%%%%%%%%
%% Introduction
%%%%%%%%%%%%%%%%%%%%%%%%
\iffalse
\section*{Comments from CWR}
\href{https://docs.google.com/document/d/17jX85aERNViOyQ-qqlpR6yGKUV4-P3GJ/edit}{Link to comments}
\fi 

\section{Introduction} \label{sec:intro}

The Sunyaev-Zel'dovich (SZ) effect \citep{sunyaev70b,sunyaev80} is one of the largest sources of secondary anisotropies in the cosmic microwave background (CMB) and enables powerful probes of astrophysics and cosmology \citep[e.g.,][]{birkinshaw99, carlstrom02}.
The SZ effect occurs when free electrons, such as those found in the hot intracluster medium of galaxy clusters, Compton scatter CMB photons. 
The SZ effect can be subdivided into two main subclasses. 
The first is the thermal SZ (tSZ) effect, which occurs due to an energy transfer from the hot electrons to the CMB photons during inverse Compton scattering. 
The net effect is to slightly distort the CMB blackbody spectrum by upscattering some CMB photons to higher frequencies.
The second is the kinematic SZ (kSZ) effect, which is the Doppler shift induced by the bulk velocity of the free electrons and modifies the observed temperature of the CMB blackbody spectrum. 
The kSZ effect is the focus of this work. 

Observations of the kSZ effect can be used to constrain both cosmological and astrophysical parameters \citep[e.g.,][]{lahav91, haehnelt96, battaglia13, ma13, bianchini16, alonso16, smith17,  battaglia17, alvarez21, raghunathan23, raghunathan24}. 
Measuring the kSZ effect is challenging due to the kSZ effect's faintness and spectral degeneracy with the CMB temperature fluctuations \citep[]{shaw12}. 
However, the kSZ effect is also one of a handful of observables that can directly probe the large-scale velocity field. 
The pairwise kSZ effect looks at the average relative velocity of two clusters as a function of their physical separation, which on average, these clusters should be falling towards one another due to gravity. 
This is one of several approaches that can be used to measure the kSZ signal.
The first detection of the pairwise kSZ effect \citep{hand12} was made using high-resolution CMB data from the Atacama Cosmology Telescope (ACT) \citep{ACT} in conjunction with the Baryon Oscillation Spectroscopic Survey (BOSS) data release 9 spectroscopic galaxy catalog \citep{BOSSdr9}. 
A subsequent pairwise kSZ analysis was performed using ACT and the Sloan Digital Sky Survey (SDSS) DR15 spectroscopic data set \citep{calafut21}, which measured the pairwise kSZ signal to 5.4$\sigma$. This is the highest significance measured so far due to the use of spectroscopic catalogs.
Measurements of the kSZ signal with the pairwise statistic have also been carried out using photometric catalogs, notably by the South Pole Telescope \citep{padin08, carlstrom11} and the Dark Energy Survey (DES) \citep{flaugher15} collaborations, which reported detections of the pairwise kSZ at the level of around $4\sigma$ \citep{soergel16,schiappucci22}.
Other detections of the pairwise kSZ effect with data from the photometric DESI Legacy Imaging Surveys and \planck data have been reported by \cite{chen21,li24}.

Several other methods have been suggested to detect the kSZ signal. 
Among these are the velocity-weighted stacking approach \citep{li14,schaan16,schaan21, hadzhiyska24}, kSZ tomographic analyses \citep{ho09,smith18}, and estimators based on projected fields \cite{hill16,kusiak21,patki23}. 
Additional strategies include using a velocity reconstruction method \citep{deutsch18}, angular redshift fluctuations \cite{Chaves_Montero_2020}, performing cross-correlations with 21 cm and other line-intensity mapping data sets \citep{sato-polito21,li19}, and focusing on local measurements for individual clusters \citep{sayers13}.
The variety of techniques explored to detect and analyze the kSZ effect offers the potential for an enhanced understanding of cosmic velocities and structures.

In this work, we forecast the cosmological constraining power of pairwise kSZ measurements with the next-generation CMB-S4 \citep{cmbs4} experiment and upcoming large-scale structure surveys, like the Vera C. Rubin Observatory’s Legacy Survey of Space and Time (LSST) \citep{ivezic19} and the 4-metre Multi-Object Spectroscopic Telescope (4MOST) \cite{4most} with the eROSITA cluster follow-up survey \citep{erosita19}. After near-term stage-3 experiments (like ACT and SPT), CMB-S4 is the next stage for ground-based CMB experiments and will employ both large and small aperture telescopes in order to probe small and large angular scales.
In this work, we restrict ourselves to the wide-area survey of CMB-S4, which will cover $\sim 50\%$ of the sky after removing the high emission area of our galaxy (See Fig.~1 of \citep{raghunathan21}). 
We will use a spectroscopic galaxy cluster catalog, since photometric redshifts dilute the pairwise signal by a factor of $2$, as shown in \citep{schiappucci22}.
We improve the cosmological constraints from the pairwise kSZ measurement by calibrating the optical depth measurements to the tSZ signal of the cluster catalog. Finally, we consider the effects of scatter in the mass-richness scaling relation and the impact of mis-centering biases that can occur when dealing with the measurements from optical galaxy cluster surveys.

\iffalse
The paper is organized as follows.
 In Section \ref{sec:theory}, we briefly describe the theory behind the kSZ and tSZ effects and the pairwise velocity technique.
\sr{Section III is not mentioned here.}
In Section \ref{sec:analysis}, we describe the method used for the pairwise kSZ measurement.
In Section \ref{sec:ksz_measurement}, we give the forecast results for the pairwise kSZ measurement and its significance level over different redshift bins. We will also make a Fisher forecast on the cosmological constraints.
 Finally, we briefly summarize our results in Section \ref{sec:conclusions}, and discuss the main implications for future analyses.
 \sr{I think this whole paragraph is not required.}
 \fi

%%%%%%%%%%%%%%%%%%%%%%%%
%% Theoretical Background
%%%%%%%%%%%%%%%%%%%%%%%%
\section{Theoretical Background} \label{sec:theory}

\subsection{The kinematic Sunyaev-Zel'dovich effect}

The kSZ effect yields a small change in the CMB blackbody temperate, $\Delta T_{\rm kSZ}$, which is proportional to the product of the free electron number density and line-of-sight velocity. 
The dependence of the kSZ signal on the bulk velocity of the ionized gas in the cluster offers a unique opportunity to probe the cosmological velocity field \citep[]{bernardeau02}.
The optical depth of a galaxy cluster to CMB photons is low (typically $\tau_e \lesssim 0.01$), so we can use the single-scattering limit. 
In this limit, the change in the CMB temperature due to the kSZ effect is
\begin{equation}
    \frac{\Delta T_{\rm kSZ}}{T_{\rm CMB}} = - \sigma_T \int d\ell\, n_{e} \frac{\rver \cdot \mathbf{v}}{c} \simeq - \tau_{e} \frac{v_{\text{los}}}{c}, 
\label{eq:ksz}
\end{equation}
where $c$ is the speed of light, $\sigma_T$ is the Thomson cross section, $n_{e}$ is the number density of electrons, and $\tau_{e}$ is the Thomson optical depth for CMB photons traversing the galaxy cluster \citep{sunyaev80}. 
The unaltered CMB temperature is $T_{\rm CMB}$, while $\bf{v}$ is the cluster velocity and $\hat{\mathbf{r}}$ a unit vector in the line of sight.

\subsection{The pairwise kSZ signal}

On scales smaller than the homogeneity scale ($\sim350$ Mpc \cite{yadav10}), pairs of galaxy clusters are expected to fall towards one another due to their mutual gravitational pull. 
This infall will produce a small dipole pattern in the CMB temperature anisotropy at the clusters' positions due to the kSZ effect (e.g., \citep{diaferio00}), which is known as the pairwise kSZ signal. 
The small pairwise kSZ signal can be detected by stacking many such pairs of clusters. 
The average pairwise kSZ amplitude $T_{\rm pkSZ}(r)$ for such a stack of clusters at a comoving separation $r$ can be related to the mean pairwise velocity $v_{12}(r)$ of the clusters:
\begin{equation}
    T_{\rm pkSZ}(r) \equiv \bar{\tau}_e \frac{v_{12}(r)}{c} T_{\rm CMB},
\label{eq:TpkSZ}
\end{equation}
where $\Bar{\tau}_e$ is the average optical depth of the sample. This equation requires two approximations: (1) the internal motion of the cluster's gas is negligible; and (2) the optical depth and velocity of the clusters are uncorrelated \citep{soergel18}.
We adopt a sign convention so that clusters falling towards one another will have a negative relative velocity $v_{12}(r)$ and negative $T_{\rm pkSZ}$ signal.

\begin{figure}[tbp]
\centering
\includegraphics[width=\linewidth]{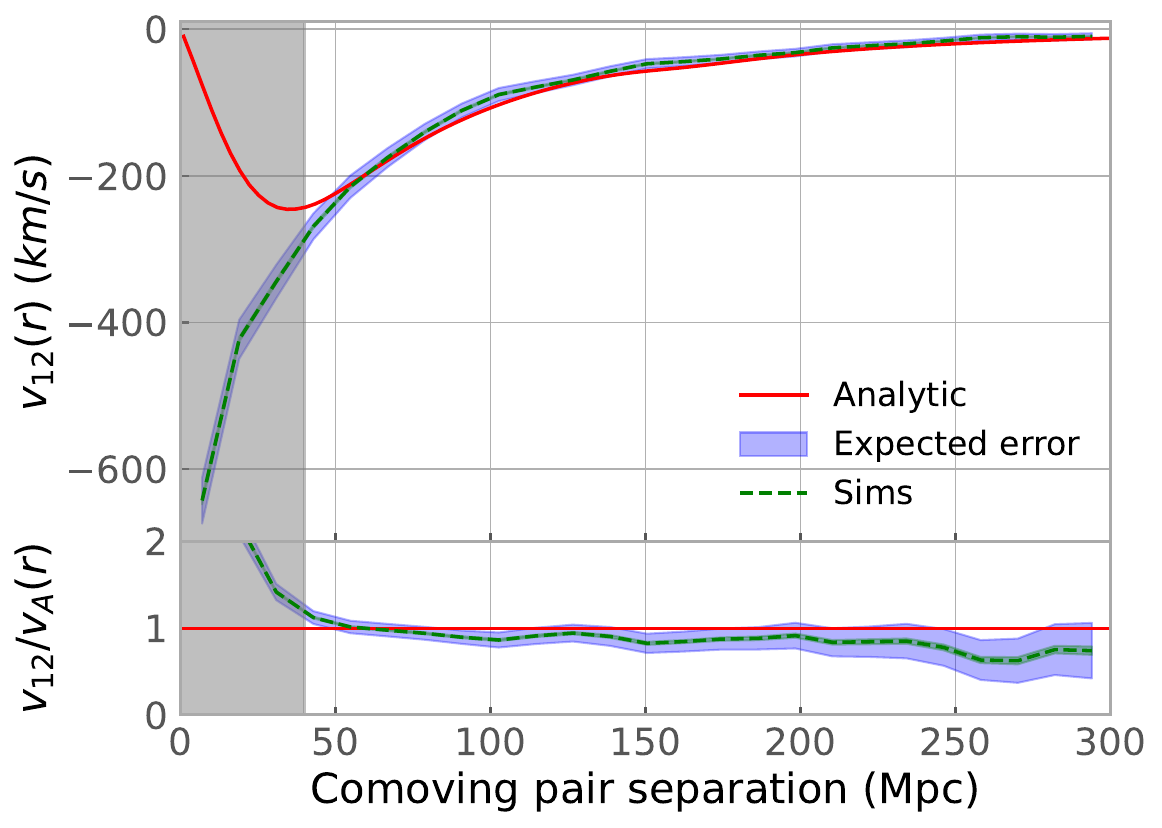}
\caption{Top: Comparison between the analytical model (solid red line) of the mean pairwise velocity $v_{12}(r)$ compared to one obtained through simulations described in Section~\ref{sec:sims} (dashed green line). The shaded blue area shows the expected error bars from the pairwise kSZ reconstruction for CMB-S4. Bottom: Quotient of the mean pairwise velocity measured from simulations with respect to the analytical model $v_A(r)$. The shaded gray region indicates the non-linear regime ($r < 40$ Mpc) where the analytical model is not reliable.}
    \label{fig:v12model}
\end{figure}

The mean pairwise velocity of haloes $v_{12}(r)$ separated by comoving distance $r = |\vec{r}_2 - \vec{r}_1|$ can be analytically modeled in linear theory for a specific cosmology and theory of gravity in terms of the two-point matter correlation function $\xi (r)$ as \citep{juszkiewicz98,sheth00,bhattacharya07} 

\begin{equation}
\begin{aligned}
v_{12}(r,a) \approx -\frac{2}{3} a H(a) f(a) r \frac{b\bar{\xi}(r)}{1+b^2\xi(r)},
\end{aligned}
\label{eq:v12}
\end{equation}
where $a$ is the scale factor, $H(a)$ is the Hubble parameter, $f(a)\equiv {\rm d}\ln D/{\rm d}\ln a$ is the growth rate (with $D$ being the linear growth factor), $b$ the mass-averaged halo bias, and $\Bar{\xi}$ indicates the average of $\xi(r)$ over a comoving sphere of radius $r$. Fig.~\ref{fig:v12model} compares the theoretical approximation of the pairwise velocity with a measurement obtained from the AGORA simulations, described in Sec.~\ref{sec:sims}.

Equations (\ref{eq:TpkSZ}) and (\ref{eq:v12}) highlight how measurements of the pairwise kSZ are sensitive to a combination of both cluster astrophysics, through the optical depth $\bar{\tau}_e$ and halo bias $b$, and cosmology through the Hubble parameter $H(a)$, the growth rate $f$, and the two-point matter correlation function $\xi (r)$. 
In particular, the dependence on the growth rate $f$ and matter correlation function $\xi(r)$ makes measurements of the pairwise kSZ signal sensitive to $f \sigma_8^2$, thus providing complementary information to other cosmological probes such as redshift-space distortions, primarily sensitive to $f \sigma_8$ \citep{percival09}. Therefore, if we break the degeneracy with cluster astrophysics through alternative methods, we can use the pairwise kSZ signal to access information of the velocity field (which on large scales is entirely dependent on the growth of the density field) and probe dark energy and modifications of gravity \citep{bhattacharya07,kosowsky09,keisler13,mueller15, ma15}.

\subsection{The thermal Sunyaev-Zel'dovich effect}

As noted in the introduction, the kSZ effect is the fainter of the two SZ effects associated with a galaxy cluster. 
The larger signal is the tSZ effect, which is due to inverse Compton scattering between CMB photons and the hot electrons of the intracluster medium \citep{sunyaev70,carlstrom02}. 
The magnitude of the tSZ effect for a galaxy cluster can be quantified by the Compton-$y$ parameter \citep{sunyaev72}, and expressed as an integral over the line of sight:
\begin{equation}
y(\rver_i) = \sigma_T \int {\rm d}\ell\, n_e \frac{k_B T_e}{m_e c^2}.
\end{equation}
Here $n_e$ is the free electron density, and $T_e$ is the electron temperature.
The physical constants are the speed of light $c$, Boltzmann constant, $k_B$ and electron mass $m_e$. 
Importantly, both the tSZ and kSZ effects depend on the integrated electron number density, albeit with different weightings (temperature versus velocity). 
Thus the measured tSZ signal can potentially be used to calibrate the optical depth dependence of the kSZ signal (see \citep{battaglia16b, soergel18, vavagiakis21}). 
We return to this in Sec.~\ref{sec:ytau}.

%%%%%%%%%%%%%%%%%%%%%%%%
%% Simulations 
%%%%%%%%%%%%%%%%%%%%%%%%
\section{Simulations} \label{sec:sims}

\subsection{Simulated CMB maps}

We use realistic realizations of the millimeter wavelength sky from The Agora: Multi-Component Simulation for Cross-Survey Science \citep{omori22}, which leverages the N-body simulations of the MultiDark \Planck 2 Synthetic Skies suite \citep{klypin16}.
This simulation is based on \Planck's \lcdm model \citep{planck18-6} with the following cosmological parameters: $H_0=67.77\,{\rm km\, s}^{-1}\,{\rm Mpc}^{-1}$, $\Omega_m=0.307115$, $\Omega_b=0.048206$, $\sigma_8=0.8228$, $n_s=0.96$.\footnote{The cosmological parameters listed are the Hubble parameter, matter density, baryonic matter density, current root mean square (rms) of the linear matter fluctuations on scales of $8h^{-1} {\rm Mpc}$, and the spectral index of the primordial scalar fluctuations,  respectively.}
The simulated skies are generated by pasting astrophysical effects onto the dark matter haloes, obtained through a friends-of-friends algorithm applied to the MultiDark \Planck 2 $N$-body simulation \citep{klypin16}. 
The astrophysical modeling in the simulation has been calibrated using observational data and external hydrodynamical simulations.

Outlined below are the main components of the simulated microwave sky. 
\begin{itemize} 
\item The dark matter density field is used to gravitationally lens the CMB sky.
\item The tSZ signal from each dark matter halo is added based on the gas profile as described in \citep{mead20}. The electron thermal pressure profile was calibrated on the hydrodynamical BAHAMAS simulations suite \citep{mccarthy16}.
\item The kSZ effect is added in a similar way as the tSZ effect. The same \citep{mead20} gas profile is used to estimate the electron number density, which is multiplied by the line-of-sight velocity to obtain the kSZ signal from the halo. 
\item The simulations also include correlated signals from dusty star-forming galaxies, also called the cosmic infrared background (CIB). 
\item Radio galaxies, which appear as point sources in the simulation maps, have been masked and are not included in the measurements.
%The cosmic infrared background (CIB) from dust enshrouded galaxies is simulated by first assigning star formation rate and stellar mass to each individual halo using the UniverseMachine code \citep{behroozi19}. With that information, the bolometric infrared luminosity is inferred from \citep{kennicutt98}, and then converted to flux density assuming the shape of the spectral energy distribution to be a modified blackbody. \sr{Not sure if these details are important. But if they are important, then we probably need more details. So I would be inclined to just explain tSZ and kSZ, and then say that the simulations also include correlated signals from dusty star forming galaxies and radio galaxies.}
%\item The radio galaxies, which appear as point sources in the simulation maps, have been masked and are not included into the measurements.
\end{itemize}

We convolve the simulated maps by the expected instrumental beams for each CMB-S4 frequency band and add the expected white and 1/f noise levels to each map. Following \citep{raghunathan21}, the assumed map noise has the form

\begin{equation}
    N_\ell = \Delta_T^2 \bigg[1 + \bigg(\frac{\ell}{\ell_{knee}} \bigg)^{-\alpha_{knee}} \bigg]. \label{eq:noise}
\end{equation}
Here $\Delta_T$ is the detector noise level, $\ell_{knee}$ and $\alpha_{knee}$ are used to estimate the atmospheric $1/f$ noise, and $\ell$ is the spherical harmonic index. 
The assumed values of these parameters for each of the three relevant frequency maps are listed in Table~\ref{tab:CMB-S4-specs}. 

\begin{table*}
    \centering
    \begin{tabular}{cccc}
        \hline
        \hline
        Frequency & Beam FWHM  & Noise level  & 1/f noise  \\
        (GHz) & ($^\prime$) & ($\mu K-\text{arcmin}$) & ($\ell_{knee}$, $\alpha_{knee}$)\\
        \hline
        90 & 2.5 & 2.0 & (2154, 3.5) \\
        150 & 1.6 & 2.0 & (4364, 3.5) \\
        220 & 1.1 & 6.9 & (7334, 3.5) \\ 
        270 & 1.0 & 16.7 &  (7308, 3.5) \\
        \hline
        \hline
    \end{tabular}
    \caption{The assumed instrumental beam FWHM (column 1) and map noise specifications for each CMB-S4 frequency band (columns 2 and 3). 
    Following \citep{raghunathan21} and Eqn.~\ref{eq:noise}, the map noise is described by a white noise level ($\Delta_T$) and 1/f noise parameters ($\ell_{knee}$, $\alpha_{knee}$).}
    %The noise level and the 1/f noise gives the information on the $1/f$ noise that is described by equation \ref{eq:noise}.}
    \label{tab:CMB-S4-specs}
\end{table*}

\subsection{Mock cluster catalog}

Mock cluster catalogs are a key ingredient to the pairwise kSZ forecasts of this work. 
We expect the cluster catalogs for the CMB-S4 measurement will come from an optical survey, such as LSST or 4MOST. 
Existing optical cluster finders, such as \redmapper{} \citep{rykoff14}, identify galaxy clusters as overdensities of red sequence galaxies. The number of red sequence galaxies in clusters (often referred to as richness $\lambda$) is correlated with halo mass and is commonly used as the mass tracer of galaxy clusters. 

This approach to cluster detection can suffer from systematic uncertainties, most significantly projection effects due to the chance alignment of structures along the line of sight \citep{DES_cluster, To+, tomommi+}.
Fully simulating these effects is difficult and beyond the scope of this work, as the magnitude is sensitive to galaxy colors, luminosity, and their environmental dependence \citep{To+, Cardinal}.
However, \citep{DES22} demonstrates that an approximate treatment, based on counting simulated galaxies within a cylinder along the line of sight, can capture the effect of projection reasonably well. 
We follow the same cylinder selection method in this work. 
However, we use a different projection length of $\pm30$ Mpc/h since we find this better reproduces the observed number counts $N(\lambda)$ as a function of richness in the DES Year 1 data \citep{DES_cluster}. 

Within the CMB-S4 footprint, the simulated cluster sample contains $N=654,928$ clusters in the redshift range $z \in [0.2, 1.2]$, with an optical richness estimate $\lambda > 20$. This corresponds to the redshift and richness range for which the future LSST will provide a complete cluster sample \citep{ivezic19}. 
The photometric redshift estimates from LSST reduce the signal-to-noise ratio by half, as shown in previous works \citep{schiappucci22}. Therefore, we assume 4MOST will provide the spectroscopic redshifts for the cluster catalog. 

%%%%%%%%%%%%%%%%%%%%%%%%
%% Methods
%%%%%%%%%%%%%%%%%%%%%%%%
\section{Analysis Methodology} \label{sec:analysis}

\subsection{Pairwise kSZ estimator} \label{sec:TpkSZ}
Following previous works \citep{hand12,calafut21,schiappucci22}, we implement the pairwise kSZ estimator $\hat{T}_{\rm pkSZ}(r)$ introduced by \citep[]{ferreira99}, which has the form

\begin{equation}
    \hat{T}_{\rm pkSZ}(r) = - \frac{\sum_{i<j,r} [T(\rver_i) - T(\rver_j)] \; c_{ij} }{\sum_{i<j,r} c^2_{ij}}. \label{eq:pkSZestimate}
\end{equation}
This estimator scales the CMB temperature difference at the location of two clusters by a geometrical factor,  $c_{ij} = \hat{\bf r}_{ij} \cdot (\hat{\bf r}_i + \hat{\bf r}_j)/2$, to account for the projection of the pair separation $\hat{\bf r}_{ij} = \hat{\bf r}_i - \hat{\bf r}_j$ onto the line of sight. 
The temperature difference will depend on the difference between the kSZ signal in the two clusters, which in turn depends on the line of sight projection of the relative velocity between them. 
 
We reconstruct the pairwise kSZ signal in eight equal and linearly spaced bins between comoving pair separation $r$ of 10 and 300 Mpc, with a bin width of $\Delta r = 36.25$ Mpc.

\subsection{Map filtering and temperature extraction} 
\label{sec:Textraction}

To improve signal-to-noise, we use a constrained internal linear combination (cILC) approach \citep{remazeilles11gilc} to combine the simulated multi-frequency maps for CMB-S4. 
Combining the different frequency bands in this way will lead to a CMB+kSZ map, while minimizing noise and the influence from frequency-dependent foregrounds, and simultaneously explicitly removing the tSZ signal. We combine data from different frequency channels using the cILC technique in harmonic space as

\begin{equation}
    S_\ell = \sum_{i=1}^{N_{\nu}} \omega_{\ell}^i M_{\ell}^i,
\end{equation}
where $S$ is the signal we want to retrieve from the frequency maps $M^i$, $N_{\nu}$ is the total number of frequency channels, and $\omega^i$ is the $\ell$-dependent weights for each frequency channel. These weights are tuned to produce a minimum-variance map along with nulling the contribution of the tSZ signal, where the specific frequency response is known, as 

\begin{equation}
    \omega_\ell = \mathcal{C}_\ell^{-1} \mathcal{F} (\mathcal{F}^{\dagger} \mathcal{C}_\ell^{-1} \mathcal{F})^{-1} N,
\end{equation}
where the matrix $\mathcal{C}_\ell$ has dimension $N_{\nu} \times N_{\nu}$ and contains the covariance between simulated maps, $\mathcal{F}$ contains the frequency response vector of the desired signal and the foregrounds, and $N$ is used to extract the desired signal as described in \citep{remazeilles11gilc}. 

To extract the CMB+kSZ temperature from the cILC combined map for Sec.~\ref{sec:pksz_measurement}, we use an aperture photometry filter on the positions of clusters, which is written in map space as

\begin{equation}
    \Psi (\theta) = \frac{1}{\pi \theta_r^2} \times
    \begin{cases}
        1 & 0 < \theta < \theta_r , \\
        -1 & \theta_r < \theta < \sqrt{2} \theta_r , \\
        0 & \text{elsewhere} ,
    \end{cases}
\label{eq_ap_photo}
\end{equation}
where $\theta_r$ is the characteristic filter scale. 
The aperture photometry acts as a high pass filter for scales that are larger than the filter scale by subtracting the average temperature in the outer ring from the average temperature inside the disc of radius $\theta_r$. 
In contrast to matched filter techniques used in \citep{soergel16,schiappucci22}, this approach does not assume a specific model for the cluster profile, but it requires that the cluster is contained within the characteristic filter scale $\theta_r$ to avoid biases in the temperature estimation $\hat{T}_0(\nver_i)$. 

\begin{figure}[!htbp]
\centering
\includegraphics[width=\linewidth]{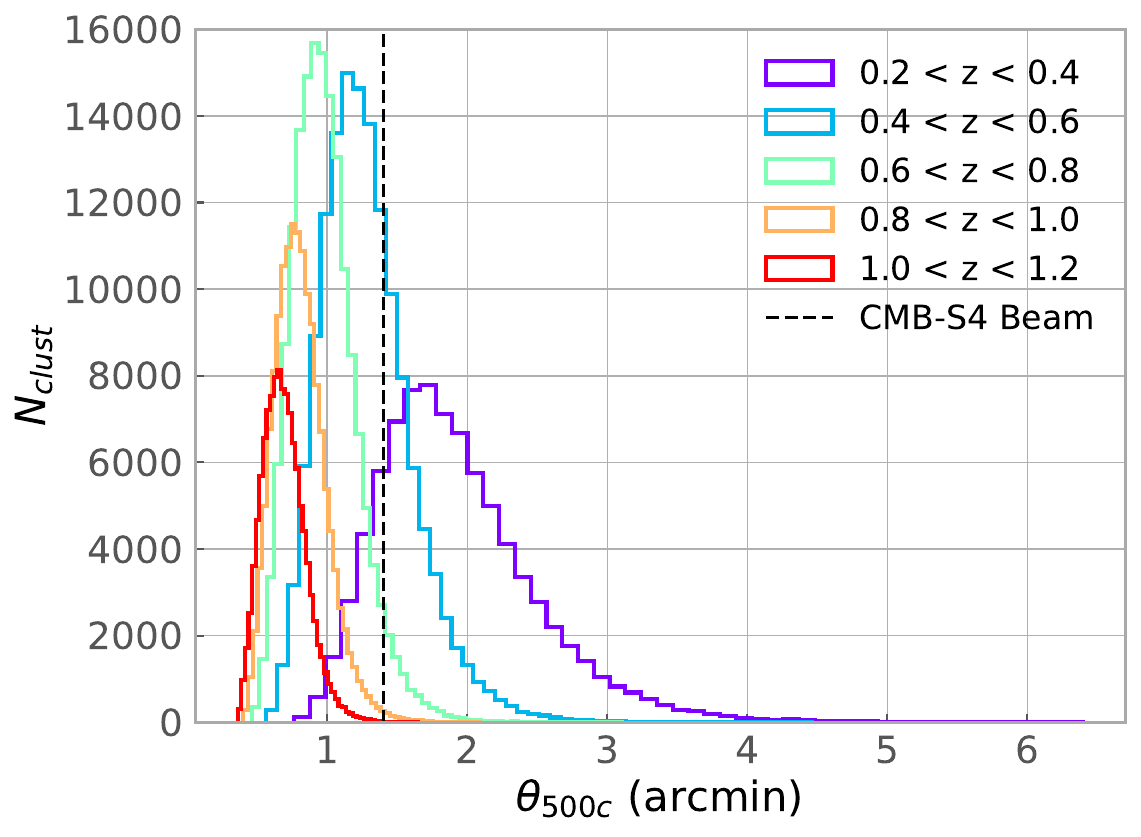}
\caption{Histogram of $\theta_{500c}$ of the clusters for the aperture photometry in different redshift bins. We see how the peak of the angular size of clusters decreases with redshift. The black dashed line is the effective FWHM beam of CMB-S4.}
    \label{fig:theta500hist}
\end{figure}

In this work, we use the radius $\theta_{500c}$ which is defined as the sphere within which the cluster mass is 500 times the critical density of the Universe at the cluster redshift. This aperture could miss some of the gas in the cluster, thus a model of what fraction is covered within that aperture will be necessary for future analyses. The distribution of $\theta_{500c}$ in our cluster catalog is shown in Fig.~\ref{fig:theta500hist}. 

\subsection{Redshift-dependent foregrounds} \label{sec:estimator} 

Over an extended redshift range, the redshift evolution of foregrounds (like the CIB) can introduce a redshift-dependent bias in the estimated temperatures at the positions of clusters \citep{schiappucci22}. To mitigate this, we estimate the mean measured temperature as a function of redshift and subtract this mean temperature from the estimated temperatures from the aperture photometry $\hat{T}_0(\nver_i)$, as

\begin{equation}
T\left(\hat{\mathbf{n}}_{i}\right)=\hat{T}_{0}\left(\hat{\mathbf{n}}_{i}\right)-\frac{\sum_{j} \hat{T}_{0}\left(\hat{\mathbf{n}}_{j}\right) G\left(z_{i}, z_{j}, \Sigma_{z}\right)}{\sum_{j} G\left(z_{i}, z_{j}, \Sigma_{z}\right)}.
\end{equation}
The mean measured temperature at $z_i$ is calculated from the weighted sum of contributions of clusters at redshift $z_j$ using a Gaussian kernel \mbox{$G\left(z_{i}, z_{j}, \Sigma_{z}\right) = \text{exp}[-(z_i- z_j)^2/(2\Sigma^2_z)]$}. 
As shown in \citep{hand12,soergel16}, the choice of $\Sigma_z$ does not impact the result significantly, and we set $\Sigma_z = 0.02$.  

\subsection{Covariance matrix} \label{sec:covs}

We estimate the covariance matrix of the binned pairwise kSZ measurement directly from the data using jackknife resampling \citep{escoffier16}. The jackknife resampling technique consists of measuring the pairwise kSZ signal by splitting the cluster catalogue into $N_{\rm JK}$ subsamples, removing one of them, and recomputing the pairwise kSZ amplitude from the remaining $N_{\rm JK}-1$ subsamples. This process is repeated until every subsample has been discarded once from the measurement. Then we estimate the covariance matrix as 
\begin{equation}
\hat{C}_{ij} = \frac{N_{\rm JK} - 1}{N_{\rm JK}} \sum_{\alpha = 1}^{N_{\rm JK}} (\hat{T}^{\alpha}_{i} - \bar{T}_{i}) (\hat{T}^{\alpha}_{j} - \bar{T}_{j}),
\end{equation}
where $\hat{T}^{\alpha}_{i}$ is the pairwise kSZ signal in separation bin $i$ at jackknife realization $\alpha$, of mean $\bar{T}_{i}$.

The baseline covariance matrix in this work is estimated using the jackknife resampling technique with 1000 subsamples.
The off-diagonal elements can be significant, with $15-40\%$ correlations between adjacent bins. 

%, and the correlation matrix derived from it is shown in Fig.~\ref{fig:corr_matrix}. 

We have tested the robustness of the covariance estimate to the choice of 1000 subsamples, and have found that the estimated covariance is stable. 
We show the changes in the diagonals of the covariance matrix between 1000 to 10,000 subsamples in Fig. \ref{fig:covest}.
The diagonals agree within 1.5\%, while the eigenvalues of the covariance matrices agree within 8\%. 
We conclude that 1000 subsamples are sufficient to provide a robust estimate of the covariance matrix.

\iffalse
\begin{figure}[tbp]
    \centering
    \includegraphics[width=\linewidth]{Figs/corr_matrix.pdf}
    \caption{Correlation matrix of the pairwise kSZ measurement shown in Fig.~\ref{fig:Tpksz_nozbins} calculated with 1,000 jackknife subsamples. The higher distance bins show more correlation because on average we encounter the same clusters more times.}
    \label{fig:corr_matrix}
\end{figure}
\fi

\begin{figure}[!htbp]
    \centering
    \includegraphics[width=\linewidth]{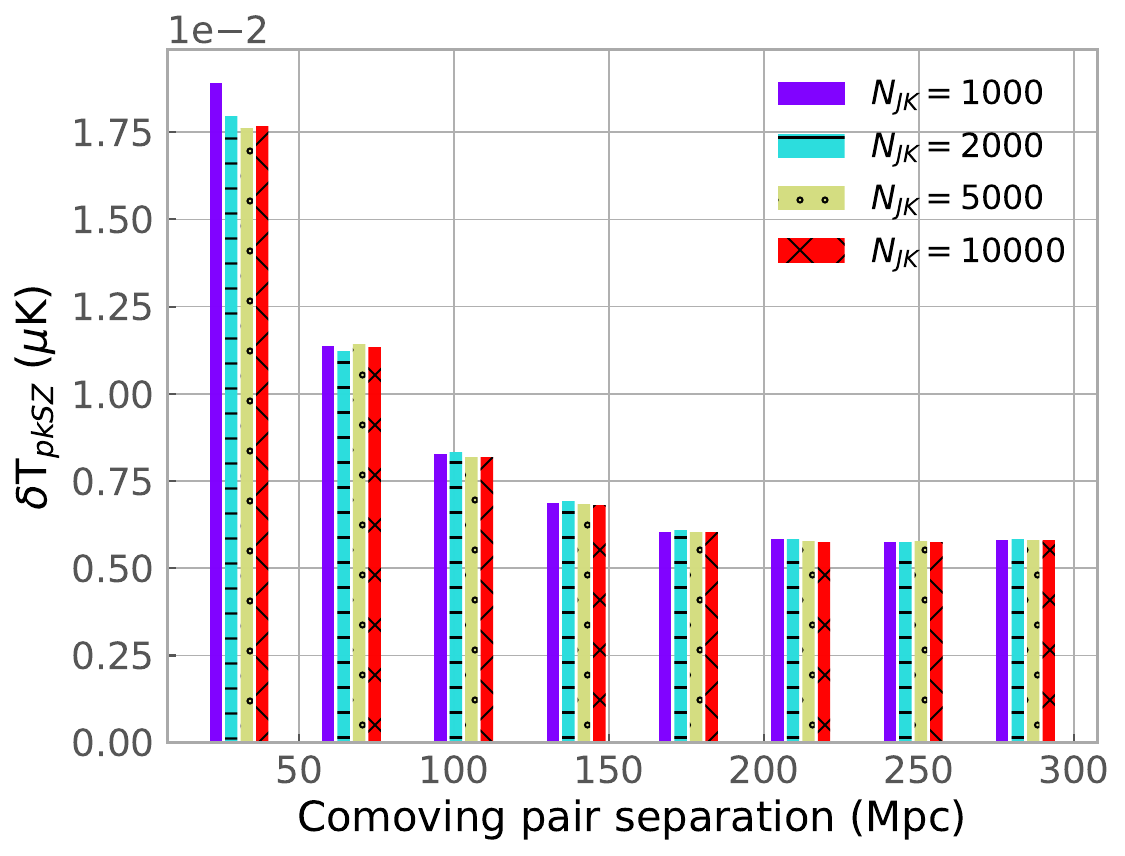}
    \caption{Estimated uncertainties for the baseline case of 1,000 subsamples (purple) compared to the recovered uncertainties for 2,000, 5,000 or 10,000 subsamples. 
    We slightly offset the results for each number of subsamples for visualization purposes. 
    The error estimate is stable across the number of subsamples.}
    \label{fig:covest}
\end{figure}

\iffalse
For the inverse of the covariance, we use the estimator

\begin{equation}
\tilde{C}^{-1} = \frac{N - N_{\rm bins} - 2}{N - 1} \hat{C}^{-1}
\end{equation}
%
where $N$ is the number of jackknife subsamples used to compute the covariance matrix $\hat{C}$, and $N_{\rm bins}$ is the number of comoving separation distance bins. This factor is needed for the covariance inversion because the empirically determined inverse covariance matrix $\hat{C}^{-1}$ is a biased estimator of the true inverse covariance matrix $C^{-1}$ as shown in \citep{hartlap06}. 
\sr{Why is this important? The factor in Eq.12 is 0.9909. How can that impact anything at all?}
\fi

%%%%%%%%%%%%%%%%%%%%%%%
%% Forecast Results
%%%%%%%%%%%%%%%%%%%%%%%
\section{Measurement forecast} \label{sec:ksz_measurement}

\subsection{Pairwise kSZ measurement} \label{sec:pksz_measurement}

\begin{figure}[!htbp]
\centering
\includegraphics[width=\linewidth]{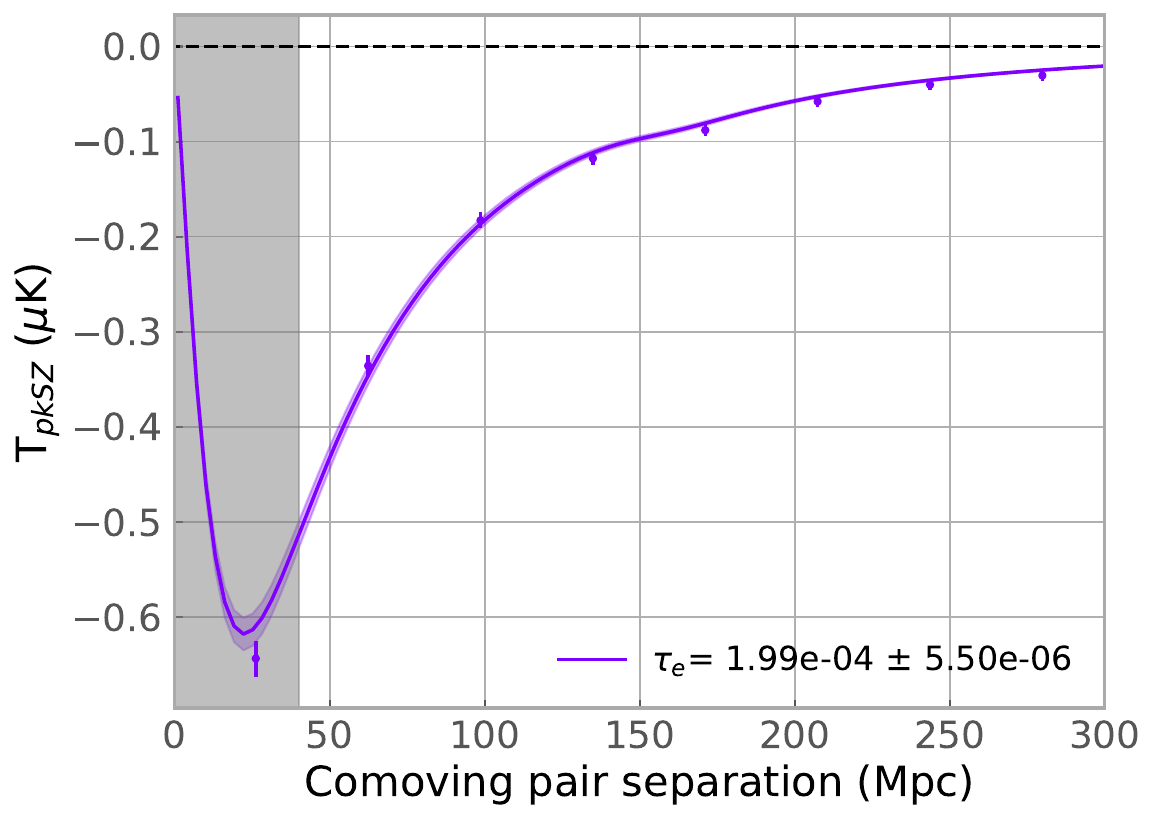}
\caption{Pairwise kSZ measurement for CMB-S4 combined with the spectroscopic redshifts from the optically selected catalog. This forecasts a measurement detection of \cmbstn$\sigma$ with respect to the null hypothesis. The grey shaded region indicates separations $r < 40$\,Mpc, where the analytical model breaks down due to the non-linear regime and is excluded from the analysis.}
    \label{fig:Tpksz_nozbins}
\end{figure}

\begin{figure}[!htbp]
\centering
\includegraphics[width=\linewidth]{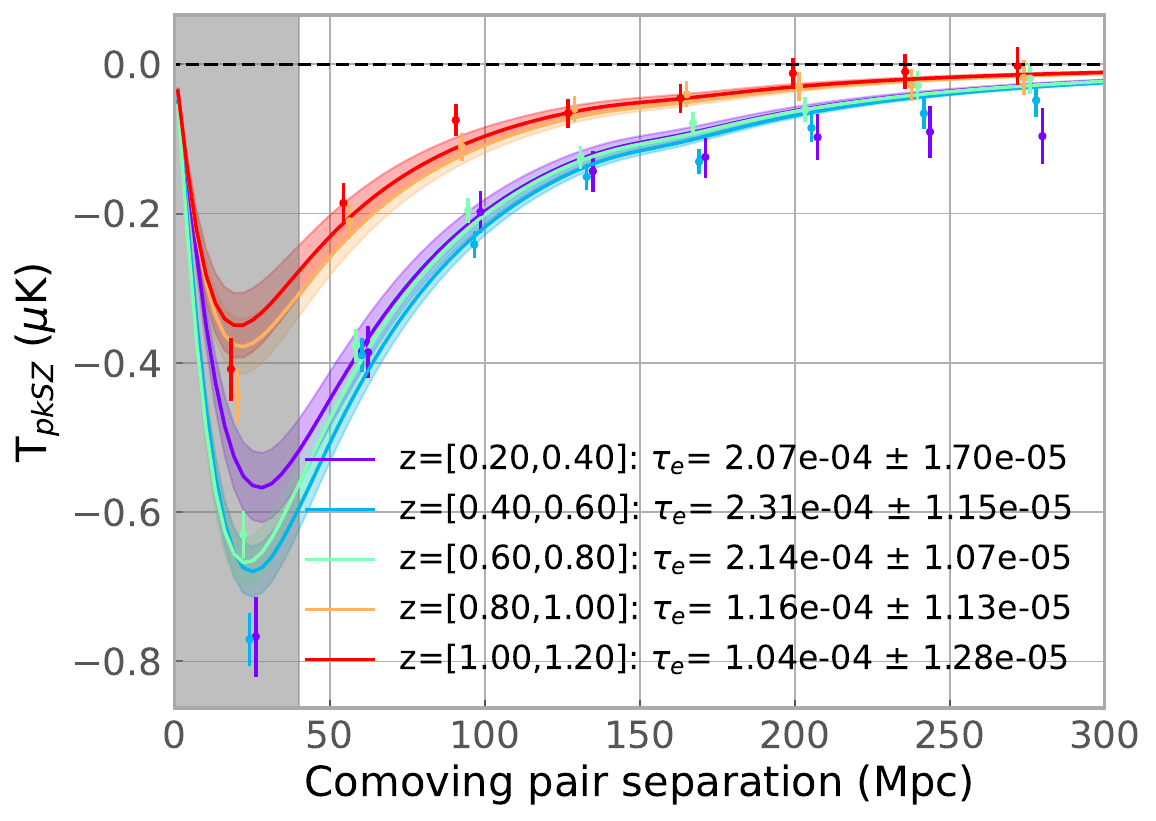}
\caption{Pairwise kSZ measurement for CMB-S4 decomposed in different redshift bins from the  optically selected catalog. This forecasts a measurement detection of \stnzbins$\sigma$ with respect to the null hypothesis for the redshift bins ([0.2,0.4), [0.4,0.6), [0.6,0.8), [0.8,1.0), [1.0,1.2]), respectively. The gray-shaded region indicates separations $r < 40$ Mpc, where the analytical model breaks down due to the non-linear regime and is excluded from the analysis.}
    \label{fig:Tpksz_zbins}
\end{figure}

We present the expected pairwise kSZ measurement for CMB-S4 in this section. 
Fig.~\ref{fig:Tpksz_nozbins} shows the results for a single redshift bin and Fig.~\ref{fig:Tpksz_zbins} decomposes the contribution from multiple redshift bins.
These results have been obtained using the cILC map described in \ref{sec:Textraction} and for optically selected clusters with richness $\lambda > 20$.

We fit the measured pairwise kSZ signal to a one-parameter model. We then compute the statistical significance of our measurement. 
The reported signal-to-noise ratio (SNR) is obtained by fixing the cosmological parameters and obtaining the best-fit $\bar{\tau}_e$ and its uncertainty by minimizing the $\chi^2$ as

\begin{equation}
\chi^2 = [\hat{T}_{\rm pkSZ} - {T}_{\rm pkSZ}(\bar{\tau}_e)]^{\dagger} \tilde{C}^{-1} [\hat{T}_{\rm pkSZ} - {T}_{\rm pkSZ}(\bar{\tau}_e)].  \label{eq:fitchisq}
\end{equation}
where $\hat{T}_{\rm pkSZ}$ is the measured signal, $\tilde{C}$ is the covariance matrix estimated using the jackknife method, and ${T}_{\rm pkSZ}(\bar{\tau}_e)$ is the model computed using Eq.(\ref{eq:TpkSZ}).
The SNR is then computed with respect to the null-signal as $\text{SNR} = \sqrt{\Delta \chi^{2}}$, with $\Delta \chi^{2} = \chi^2(\bar{\tau}_e) - \chi^2(\bar{\tau}_e = 0)$.

We forecast that the pairwise kSZ signal will be detected at a significance of $\cmbstn \,\sigma$ across all bins, and a significance of $\stnzbins\,\sigma$ for the multiple redshift bins.
These signal-to-noise numbers scale with the number of pairs of clusters, which is a function of the number of clusters in each redshift bin.
CMB-S4 in combination with future spectroscopic surveys will detect the pairwise kSZ signal at very high significance.

\iffalse
\sr{Do you understand why the tau in the first bin is several sigma off from the tau in the last bin? Is that an expected behaviour? I feel like these discussions are required. Currently, I feel that the text lacks interpretation of the results.}
\fi 

\subsection{Estimating the optical depth from the thermal SZ effect} \label{sec:ytau}

\begin{figure}[tbp]
\centering
\includegraphics[width=\linewidth]{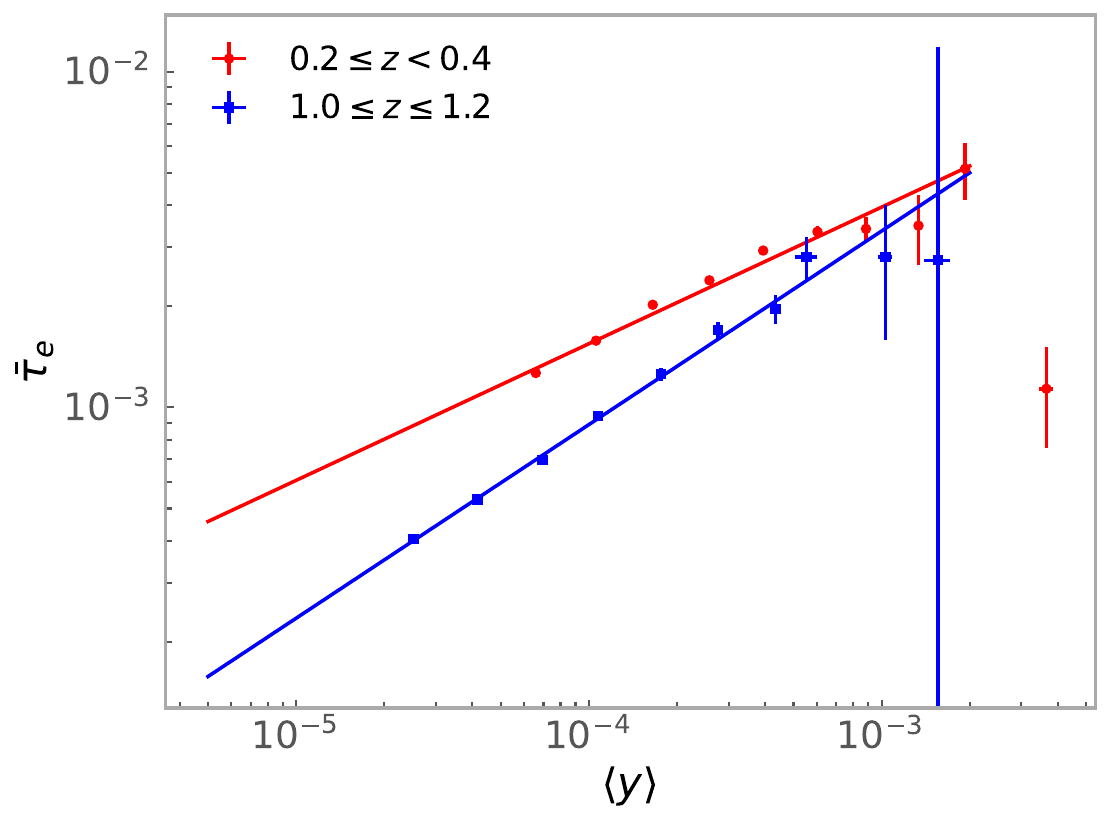}
\caption{Relationship between a cluster's Compon-y parameter and optical depth depends on redshift. 
Plotted here are the power law fits for the first ($0.2 \leq z < 0.4$, red circles) and last ($1.0 \leq z \leq 1.2$, blue squares) redshift bins. 
The fit parameters for all redshift bins can be found in Tab.~\ref{tab:tau-y_scale}. 
The expected S/N is extremely high, and there are signs that the simple power law model is an oversimplification, which is most evident for most massive clusters (largest values of Compton-y). 
Future work should explore this area more carefully.}
    \label{fig:tau_y_theory}
\end{figure}

The ability of the pairwise kSZ signal to constrain the velocity field is weakened by the degeneracy between each cluster's velocity and optical depth. 
A possible path to break this degeneracy is to use the observed tSZ signal, which also depends on the product of the number density of free electrons and electron temperature, to independently constrain the clusters' optical depth. 
Following \citep{battaglia16b, soergel18}, we assume that the relationship between the mean integrated Compton-y and mean optical depth can be represented by a power law:

\begin{equation}
\bar{\tau}_e = A \langle y \rangle^p, \label{eq:tau-y}
\end{equation}
where  $A$ and $p$ are parameters to be determined by fitting the simulations. 
We use noiseless y-maps from the Agora simulations and perform a linear least-squares fit of the relationship between the optical depth and Compton-y parameters measured by applying the aperture photometry filter described in Sec.~\ref{sec:Textraction} to each halo's lightcone. 
To allow for redshift evolution, we perform separate fits for each of the five redshift bins used in the pairwise kSZ forecasts in Fig.~\ref{fig:Tpksz_zbins}.
The best-fit parameters for each redshift bin are presented in Tab.~\ref{tab:tau-y_scale}, while Fig.~\ref{fig:tau_y_theory} shows the points and model fit for the first and last redshift bins.

\begin{table}
    \centering
    \begin{tabular}{ccc}
        \hline
        \hline
        Redshift & $A$ & $p$ \\
        \hline
        $0.2 \leq z < 0.4$ & $0.06 \pm 0.02$ & $0.41 \pm 0.04$\\
        $0.4 \leq z < 0.6$ & $0.14 \pm 0.01$ & $0.52 \pm 0.01$\\
        $0.6 \leq z < 0.8$ & $0.17 \pm 0.03$ & $0.55 \pm 0.02$ \\
        $0.8 \leq z < 1.0$ & $0.22 \pm 0.03$ & $0.57 \pm 0.01$ \\
        $1.0 \leq z \leq 1.2$ & $0.13 \pm 0.01$ & $0.58 \pm 0.01$ \\
        \hline
        \hline
    \end{tabular}
    \caption{Recovered fit parameters for the $\bar{\tau}_e - y$ relationship (see Eqn.~\ref{eq:tau-y}) in each redshift bin, as described in Sec.~\ref{sec:ytau}. 
    We find the relationship between optical depth and Compton-y to vary with redshift. The percentile uncertainty of $A$ and $m$ are 16\% and 4\%, respectively. These uncertainties are higher and lower when compared with ones obtained in \citep{battaglia16b}, which averaged 4\% and 7\%.}
    \label{tab:tau-y_scale}
\end{table}

\begin{figure}[tbp]
\centering
\includegraphics[width=\linewidth]{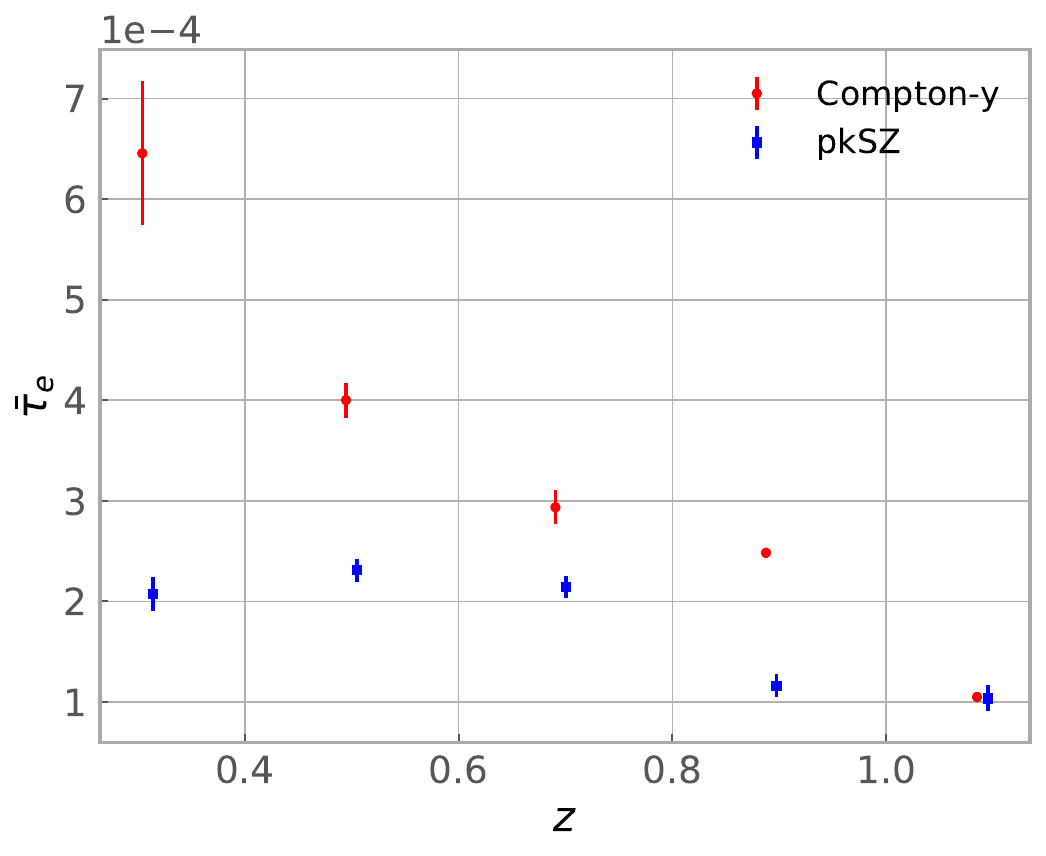}
\caption{Optical depth measurements from pairwise kSZ (blue squares) and Compton-y (red circles) for different redshift bins. The Compton-y values are shifted slightly in redshift for visualization purposes. The estimated $\bar{\tau}_e$ from equation \ref{eq:tau-y}, estimated from the parameters in Tab.~\ref{tab:tau-y_scale}, differs $(6, 9, 4, 11, 0)\sigma$ from the values inferred from the pairwise kSZ for the different redshift bins.
Future work should explore this discrepancy more carefully.
%\clr{Isn't it a big problem that you can't get simulations to agree on tau? Like shouldn't these agree by construction since we are fitting tSZ to make it match/estimate the kSZ tau? Even if the sim fit is wrong for real data, shouldn't it always work here?}
    \label{fig:taupksz_tauy}}
\end{figure}

An important caveat to this approach is that we have calibrated the Compton-y parameter to the integrated number of free electrons, but not the velocity-weighted integral of the number density that shows up in the pairwise kSZ signal. 
We examine the magnitude of this difference by comparing the mean optical depth of the simulated clusters to the velocity-weighted mean optical depth of the same clusters in each redshift bin in Fig.~\ref{fig:taupksz_tauy}. 
We find significant offsets between the two quantities, with the mean optical depth differing in each redshift bin by $(6, 9, 4, 11, 0)\sigma$, respectively, relative to the velocity-weighted values (the $\sigma$ here is only sample variance and does not include noise). 
When considering the full sample over all redshifts, the difference is $4\,\sigma$. 
%Note that the overall signal-to-noise is quite high, with the optical depth measured at $93\sigma$ by the Compton-y map. 
The discrepancy is likely due to factors like the optically selected catalog containing line-of-sight effects, which the friends-of-friends catalog does not consider. Future work should analyse why this happens and consider approaches to deal with these differences and the magnitude of the biases that would be incurred by using the mean optical depth. 

\subsection{Cosmological constraints}\label{sec:cosmo}

The pairwise kSZ is a direct measure of the infall velocity between pairs of clusters, and thus the gravitational attraction felt by these clusters. 
We consider two extensions to \lcdm{} that should lead to observable differences in this infall. 
The first is the \lcdmg{} model, where the growth index $\gamma$ is defined from the growth rate $f \approx {\Omega_m}^\gamma$. For general relativity, $\gamma \approx 0.55$, but alternate theories of gravity predict other values of $\gamma$ \citep{huterer13}.
The second extension is \wcdm{}, where the dark energy equation of state $w$ is allowed to vary from the $-1$ value of a cosmological constant. 

We use a Fisher information formalism to predict expected constraints on cosmological parameters in each model.
The covariance between two parameters $p_{\mu}$ and $p_{\nu}$, from \cite{mueller15}, is given by the Fisher matrix, which is calculated as

\begin{equation}
    F_{\mu \nu} = \frac{\partial T_{pkSZ}}{\partial p_{\mu}} \; \tilde{C}^{-1} \; \frac{\partial T_{pkSZ}}{\partial p_{\nu}},
\end{equation}
where $\tilde{C}^{-1}$ is the covariance between two pairwise kSZ bins estimated from the data in Sec.~\ref{sec:covs}. 

We determine which parameters have a higher effect on the pairwise kSZ signal by computing their log-derivatives with respect to the parameter $p_{\mu}$ being

\begin{equation}
    \nabla_{\rm log} T_{pkSZ} = \frac{\partial \rm log(T_{pkSZ})}{\partial \rm log(p_{\mu})}.
\end{equation}
The log-derivatives give an estimate of how much each parameter contributes to the pairwise kSZ signal, i.e., the higher the log-derivative, the higher the plausible constraints on that parameter. We show these derivatives in Tab.~\ref{tab:cosmo_deriv_log}, where we can see that the main cosmological parameters that contribute to the signal are $\sigma_8$, $\gamma$ and $H_0$. 

\begin{table*}[tbp]
    \centering
    \begin{tabular}{cccccccccc}
        \hline
        \hline
        Parameters & $\bar{\tau}_e $ & $\sigma_8$ & $\gamma$ & $H_0$ & $\Omega_{dm}$ & $n_s$ & $w_0$ & $\Omega_{b}$ & $w_a$ \\
        \hline
        $\nabla_{\rm log}$ & 1.00 & 0.64 & 0.21 & 0.18 & 0.09 & 0.04 & 0.02 & $9.99 \times 10^{-3}$ & $3.67 \times 10^{-6}$ \\
        \hline
        \hline
    \end{tabular}
    \caption{Derivatives on the pairwise kSZ signal as a function of cosmological parameters. The higher the value of the derivative, the better constraining power the pairwise kSZ has for that parameter. \label{tab:cosmo_deriv_log}}
\end{table*}

\begin{sidewaysfigure*}
    \centering
    \vspace{6cm}
    \subfigure[$\lcdmg$]{\includegraphics[width=0.49\textwidth]{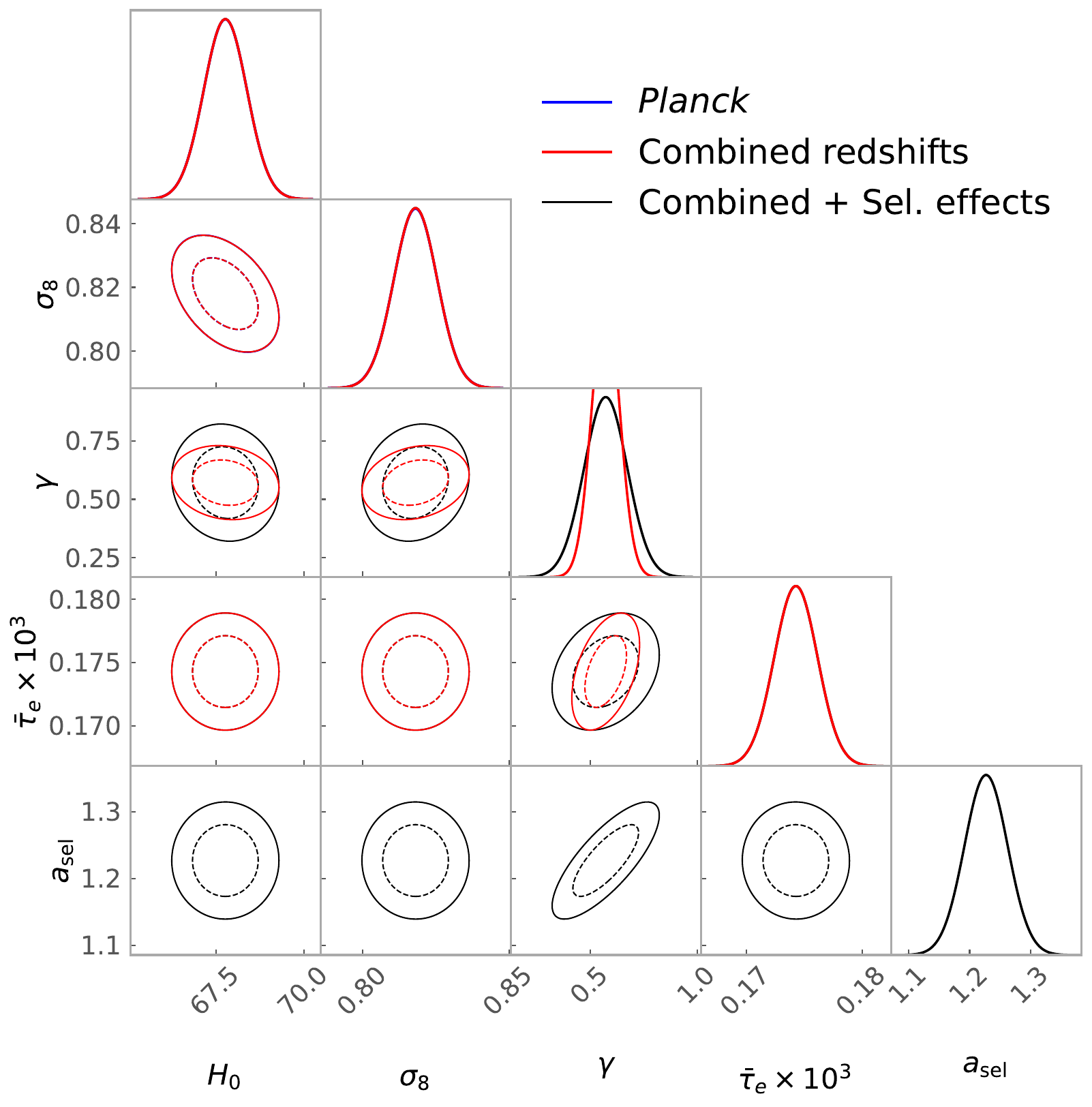} \label{fig:cosmo_4}} 
    \subfigure[$\wcdm$]{\includegraphics[width=0.49\textwidth]{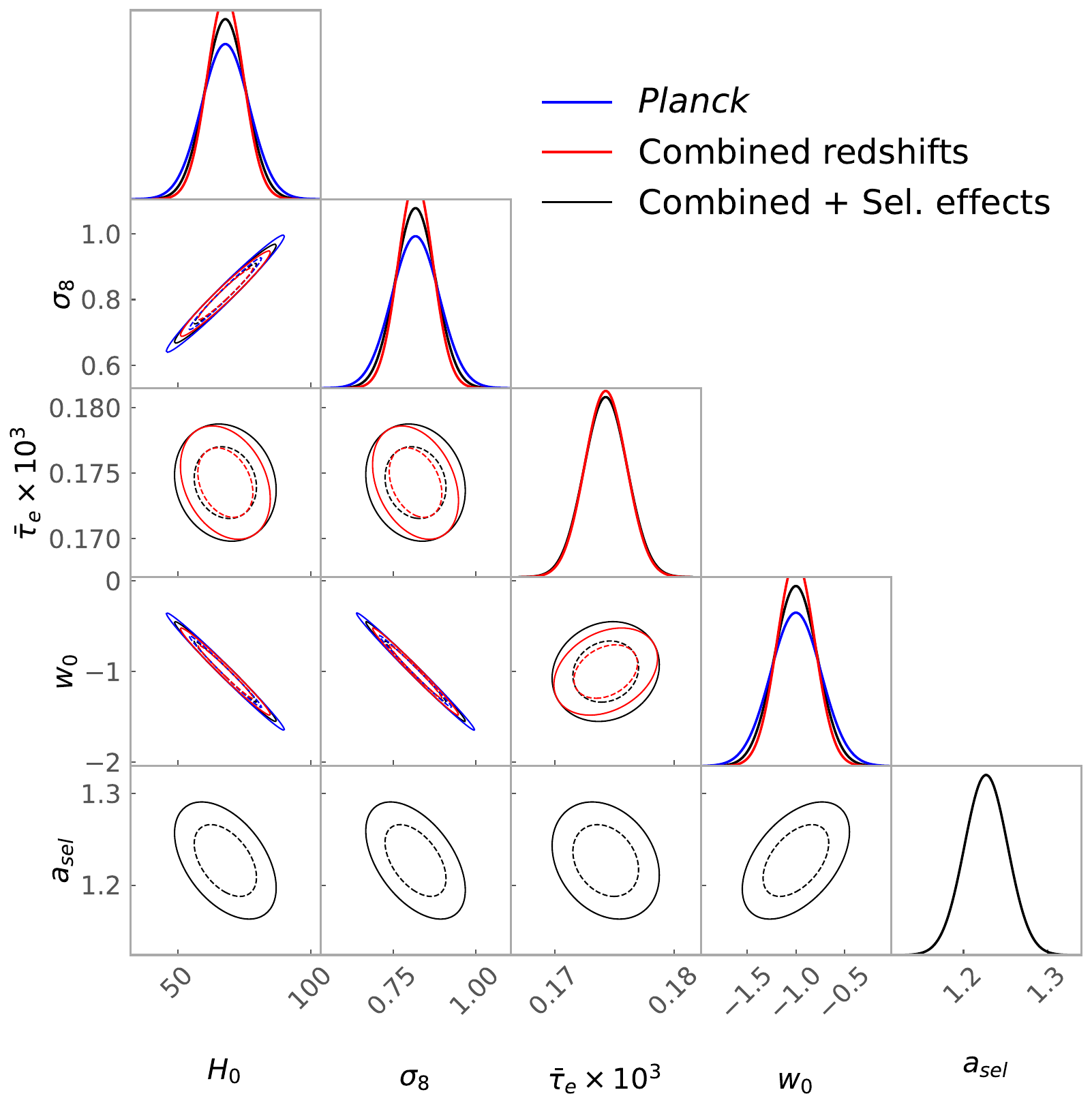} \label{fig:cosmo_omega}}
    \caption{Constraints for different cosmological models using pairwise kSZ (shown in red) with \Planck TTTEEE low$\ell$+low$E$ 2018 priors (shown in blue). The dashed lines and solid lines represent the 1$\sigma$ and 2$\sigma$ uncertainty regions, respectively. (a) cosmological parameters of the \lcdmg model, and (b) cosmological parameters of the \wcdm model. We present the parameter constraints and the improvements that pairwise kSZ measurements provide on \lcdmg and \wcdm in Tab.~\ref{tab:cosmo}. Selection effects on different cosmological models using pairwise kSZ with \Planck 2018 priors are shown in black.}
\end{sidewaysfigure*}

We show the combined constraints of the main parameters in Fig.~\ref{fig:cosmo_4} for the \lcdmg and \wcdm cosmologies. The plotted constraints include the pairwise kSZ measurement in all redshift bins, with a prior on the cluster optical depths. These constraints involve adding different redshift bins, using the mean optical depth $\bar{\tau}_e$ values and uncertainties from the $y$-map as a prior, and adding \Planck's 2018 covariance from the TTTEEE low$\ell$+low$E$ \lcdm and \wcdm MCMC chains \citep{planck18-6} as a prior to the cosmological parameters. We present the \lcdmg and \wcdm cosmological parameters in Tab.~\ref{tab:cosmo}. 

\iffalse
\sr{Why not run a LCDM-only to see if H and sigma8 can improve?}
\fi

In the  \lcdmg{} model, the growth index $\gamma$ covers the entire prior range of $0 < \gamma \le 1.0$. 
Adding the pairwise kSZ measurement tightly constrains $\gamma$, with an uncertainty of $\sigma(\gamma) = 0.02$. 
The pairwise kSZ data do not appreciably improve constraints on the other \lcdm{} parameters, with the largest fractional improvement being only 4\% for $\sigma_8$. 
There are mild degeneracies between $\gamma$ and both $\sigma_8$ and $H_0$, suggesting that some improvement might be obtained by combining the data with external probes of large-scale structure or the expansion rate. 
On the other hand, since \Planck leaves the posterior of $\gamma$ the same as the prior, we obtain a 96\% improvement. For the \wcdm cosmology, we can see how the pairwise kSZ improves on all the fiducial values from \Planck. $H_0$, $\sigma_8$ and $w_0$ are improved by 38\%, 46\%, and 42\%, respectively. 

Due to higher redshift uncertainties and/or higher CMB map noise levels, current Stage-3 measurements \citep{calafut21, schiappucci22} only detect the pairwise kSZ measurement at $\sim 5 \sigma$, compared to $36\sigma$ for CMB-S4. As a result, the Stage-3 measurements are largely unable to constrain cosmological models. For instance, using the Stage-3 uncertainties we estimate a constraint of $\gamma$ at $\sigma(\gamma) \simeq 0.5$, which is 25 times larger than the forecast for CMB-S4. Thus, while Stage-3 measurements can provide initial insights, the advanced capabilities of CMB-S4 are crucial for achieving the precision needed to significantly constrain non-standard cosmological models.

We also observe a correlation between $\bar{\tau}_e$ and all the cosmological parameters, which conveys the importance of properly constraining $\bar{\tau}_e$ through techniques like the Compton $y$ values of the clusters. We explore the full range of cosmological parameters of the \wcdm$+ \gamma + w_a$ that are not as well constrained by the pairwise kSZ in Appendix \ref{app:cosmo}.

\begin{table*}
    \centering
    \begin{tabular}{ccccc}
        \hline
        \hline
        Parameter & Fiducial & $\sigma_{\rm \Planck}$ & $\sigma_{\Planck + pkSZ}$ & Improvement over \Planck \\
        \hline
        \lcdm: \\
        \hline 
        $H_0$ & 67.77 & 0.61 & 0.57 (0.60) & 6\% (2\%) \\
        $\sigma_8$ & 0.818 & $7.42 \times 10^{-3}$ & $6.50 (7.07) \times 10^{-3}$ & 12\% (5\%)\\
        \hline
        \lcdmg: \\
        \hline 
        $H_0$ & 67.77 & 0.61 & 0.60 (0.60) & 2\% (2\%) \\
        $\sigma_8$ & 0.818 & $7.42 \times 10^{-3}$ & $7.14 (7.15) \times 10^{-3}$ & 4\% (4\%)\\
        $\gamma$ & 0.57 & 0.57 & 0.02 (0.03) & 96\% (93\%) \\
        \hline
        \wcdm: \\
        \hline
        $H_0$ & 67.77 & 8.94 & 5.51 (6.79) & 38\% (24\%) \\
        $\sigma_8$ & 0.818 & $7.19 \times 10^{-2}$ & $3.85 (5.19) \times 10^{-2}$ & 46\% (28\%) \\
        $w_0$ & -1 & 0.26 & 0.15 (0.19) & 42\% (27\%) \\
        \hline
        \hline
    \end{tabular}
    \caption{\lcdmg and \wcdm constraining power after combining the redshift bins. The fiducial values come from \Planck-TTTEEE-low$\ell$-lowE \lcdm uncertainties. The improvement represents the percentage change when adding the pairwise kSZ measurement to the \Planck cosmological constraints. The values in parenthesis are the ones when we take the selection effects into account, as described in Sec.~\ref{sec:syst}. }
    \label{tab:cosmo}
\end{table*}

\subsection{Systematics} \label{sec:syst} 

In this section, we assess the impact of systematic effects for the pairwise kSZ measurement. We consider the following sources.

\iffalse
\sr{These will also affect the y-based tau right? Why are they not considered?}
\fi

\begin{itemize}
\item {\bf Mass scatter}: In order to match the mass range from simulations, where the masses of clusters are known, to the optical cluster catalog that is selected in richness, we need a good understanding of how to obtain an accurate representation of the mass range under analysis. This is of particular interest because the analytical model in Eq.~\ref{eq:v12}, which is used to infer the optical depth of clusters, depends on the mass range of interest and changing the typical cluster mass could significantly bias this results. 
In this work, we have selected the simulation sample using the relation given in \citep{mcclintock19}. However, we need to take into account that these are estimated, and therefore a scatter in the cluster mass of the optical data can occur. To model this scatter, we draw mass errors from a normal distribution with width $\sigma_{{\rm ln}(M_{500c})} = 0.7$, which is an estimation of the scatter for a Gaussian error model in our optically selected catalog. We then use these errors to compute the pairwise kSZ signal from the simulations, obtaining an average decrease of the signal detection of $\sim 1\sigma$, which corresponds to about a $3\%$ reduction on the signal strength.

\item {\bf Mis-centering}: The measured pairwise kSZ signal can be diluted due to the fact that the position of clusters estimated from the optical survey catalog might not coincide with the location of the cluster kSZ signal. This mis-centering has a larger impact on clusters that are not fully relaxed or are merging, where the potential minimum is not located on the brightest cluster galaxy. 
To model the mis-centering, we draw RA and DEC errors from a normal distribution with width $\sigma_{\rm RA,\;DEC} = 0.5\,{\rm arcmin}$. This value corresponds to an uncertainty of \mbox{$0.2$ Mpc} at the mean redshift of the cluster catalog $\bar{z}=0.7$. This uncertainty is $3\sigma$ higher than the mean mis-centering of the optical catalog in comparison with high-resolution X-ray measurements \citep{rozo14}. This error can dilute the signal up to $5\sigma$, which corresponds to a $14\%$ decrease in the significance of the measurement.
\end{itemize}

To account for these effects, we introduce an amplitude parameter $a_{sel}$ to the pairwise kSZ measurement as 

\begin{equation}
    T_{\rm pkSZ}(r) = a_{sel} \; \bar{\tau}_e \frac{v_{12}(r)}{c} T_{\rm CMB}.
\label{eq:TpkSZ_asel}
\end{equation} 
This amplitude will constrain the amount of dilution that occurs to the pairwise kSZ measurement due to mass scatter and mis-centering of the clusters. We estimate this amplitude by calculating the ratio of the pairwise kSZ bins with respect to ones obtained after the selection effects are added to the simulations. Following \citep{diaz13}, we estimate the uncertainty as a normal approximation to the distribution of the ratio of two random variables, obtaining $a_{sel} = 1.23 \pm 0.04$.

If we include $a_{sel}$ in the Fisher formalism, using the previously obtained value as the central value, we obtain the constraints as shown in Fig.~\ref{fig:cosmo_4} for \lcdmg, and Fig.~\ref{fig:cosmo_omega} for \wcdm. We observe that $a_{sel}$ is highly correlated with all the parameters, but particularly with $\gamma$ and $w_0$. This decreases the improvements coming from the pairwise kSZ, as shown in the numbers inside the parenthesis in Tab.~\ref{tab:cosmo}. This decrease is worse for \wcdm cosmology, seeing a reduction of around $15\%$ for all the parameters. Meanwhile, for \lcdmg, there is only a 3\% reduction on the constraining power of $\gamma$. Thus, the selection effects can decrease the effectiveness of the pairwise kSZ measurement and have to be taken into account.

\iffalse
\begin{figure*}
    \centering
    \subfigure[]{\includegraphics[width=0.45\textwidth]{Figs/cosmo_params_effects_4_priors.pdf} \label{fig:cosmo_bias_4}} 
    \subfigure[]{\includegraphics[width=0.45\textwidth]{Figs/cosmo_params_effects_omega_priors.pdf} \label{fig:cosmo_bias_omega}} 
    \caption{Selection effects on different cosmological models using pairwise kSZ with \Planck 2018 priors: (a) cosmological parameters of the \lcdmg, and (b) cosmological parameters of the \wcdm. We present the parameter constraints and the improvements that pairwise kSZ provide on \lcdmg and \wcdm with the numbers inside the parenthesis in  Tab.~\ref{tab:cosmo}. Notice the different scales for each figure, particularly the \lcdmg having a smaller scale for $H_0$ and $\sigma_8$ in comparison to \wcdm.}
\end{figure*}
\fi

%%%%%%%%%%%%%%%%%%%%%%%
%% Conclusions
%%%%%%%%%%%%%%%%%%%%%%%
\section{Conclusions} \label{sec:conclusions}

We have presented forecasts for the pairwise kSZ measurement that will be achievable through the combination of the CMB-S4 survey and an LSST-like galaxy cluster catalog with spectroscopic redshifts. 
The data combination should reject the null hypothesis of no pairwise kSZ signal at $\cmbstn \,\sigma$ across all redshifts, and at a significance of $\stnzbins\sigma$ for each of five equal-width redshift bins covering the range $0.2\le z \le 1.2$. 
CMB-S4 will be able to detect the pairwise kSZ signal at very high significance. 

%\clr{Again, I don't see how this can be true, if you've constructed them to match...I can understand subgroups not matching due to poor model, but the averages should be fine by construction?}
To break the degeneracy between cosmological parameters and mean optical depth $\bar{\tau}_e$ of the clusters, we independently estimated the latter using the Compton-$y$ map.
We used a power-law form of $\tau_e = A \langle y \rangle^p$ to parameterize the relationship between the mean Compton-$y$ signal and $\bar{\tau}_e$, and calibrated this relation using noiseless simulations.
Using this calibrated relationship, the estimated $\bar{\tau}_e$ with the $y$-map differs $(6, 9, 4, 11, 0)\sigma$ relative to the values obtained with the pairwise kSZ for the different redshift bins. Meanwhile, for the full redshift range, the values disagree with each other at a $4\sigma$ level. This indicates that we need to better model the relationship between the mean Compton-y value and $\bar{\tau}_e$ with improved simulations that can tell us the optical depth of optically selected clusters. We leave this to future work. 

However, despite the disagreement, we have still shown how the pairwise kSZ signal can constrain cosmological parameters. 
Using the Fisher formalism, we produced constraints for two models: (a) \lcdmg; and (b)\wcdm. We found that $\bar{\tau}_e$ is highly correlated with $\gamma$ and $w_0$. 
This means that the better we can model and constrain $\bar{\tau}_e$, the better our constraints on modifications to cosmology will be. 
For \lcdmg, we can measure $\gamma$ to $28\sigma$, which should constrain alternatives to general relativity \citep{avila22}. 
Current Stage-3 measurements are limited by high redshift uncertainties and noise levels, achieving only a modest $5 \sigma$ detection of the pairwise kSZ effect and providing constraints on cosmological parameters like $\gamma$ that are substantially less precise than those anticipated from CMB-S4.
For \wcdm, we can improve on the \Planck cosmological parameters constraints approximately $40\%$, showing that including pairwise kSZ measurements with CMB power spectrum will help constrain alternatives to \lcdm.

We have also tested how systematics could influence the measurement of the pairwise kSZ signal. 
We tested the impact due to the scatter in the richness-mass scaling relation and the cluster mis-centering.
These effects combined can dilute the signal up to 20\% of the true amplitude value. To account for these effects, we have introduced an amplitude parameter $a_{sel}$ to the pairwise kSZ measurement. We have estimated this parameter to be $a_{sel} = 1.23 \pm 0.04$, by estimating the average of the ratios of the signal with and without the effects included. Including this additional parameter reduces the constraining power of the pairwise kSZ by 20\%.

Overall, our results show that the pairwise kSZ signal using the future CMB-S4 experiment and an LSST-like cluster catalog with spectroscopic redshifts can be a powerful tool to investigate alternatives to \lcdm cosmology, as well as shed some light on the gravitational properties of the Universe at large scales. 

\section*{Acknowledgments}

We thank Daan Meerburg and Bruce Partridge for providing useful comments on the manuscript.

SR acknowledges support by the Illinois Survey Science Fellowship from the Center for AstroPhysical Surveys at the National Center for Supercomputing Applications. 

CS acknowledges support from the Agencia Nacional de Investigaci\'on y Desarrollo (ANID) through Basal project FB210003.

The Melbourne group acknowledges support from the Australian Research Council’s Discovery Projects scheme (DP200101068). ES acknowledges support from the David Lachlan Hay Memorial Fund.

CMB-S4 is supported by the Director, Office of Science, Office of High Energy Physics of the U.S. Department of Energy under contract No. DEAC02-05CH11231; by the National Energy Research Scientific Computing Center, a DOE Office of Science User Facility under the same contract; by the Divisions of Physics and Astronomical Sciences and the Office of Polar Programs of the U.S. National Science Foundation under Mid-Scale Research Infrastructure award OPP-1935892; and by the NSF Cooperative Agreement award: AST-2240374.
Considerable additional support is provided by the many CMB-S4 team members and their institutions. 

This research used resources of the National Energy Research Scientific Computing Center, a DOE Office of Science User Facility supported by the Office of Science of the U.S. Department of Energy under contract No. DE-AC02-05CH11231. 
This research was supported by The University of Melbourne’s Research Computing Services and the Petascale Campus Initiative. 

%%%%%%%%%%%%%%%%%%%%% Ackn., bib, appendix %%%%%%%%%%%%%%%%%%%%%

\bibliographystyle{apsrev}
\bibliography{pksz_bib}

\appendix

\section{Redshift dependence of the entire cosmological parameter space}\label{app:cosmo}

\begin{sidewaysfigure*}
    \centering
    \vspace{6cm}
    \subfigure[]{\includegraphics[width=0.49\textwidth]{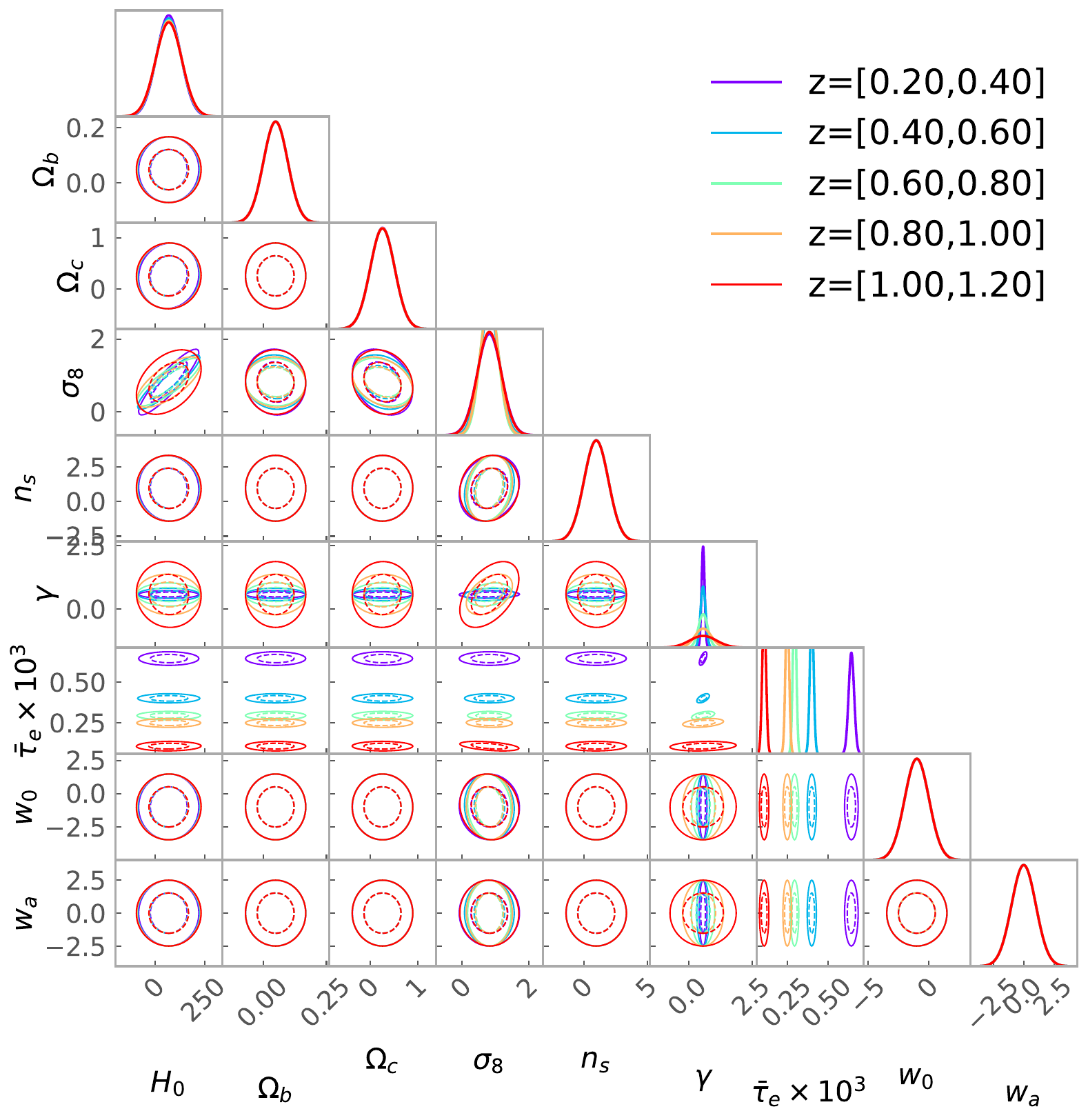} \label{fig:cosmo_zbins_all}} 
    \subfigure[]{\includegraphics[width=0.49\textwidth]{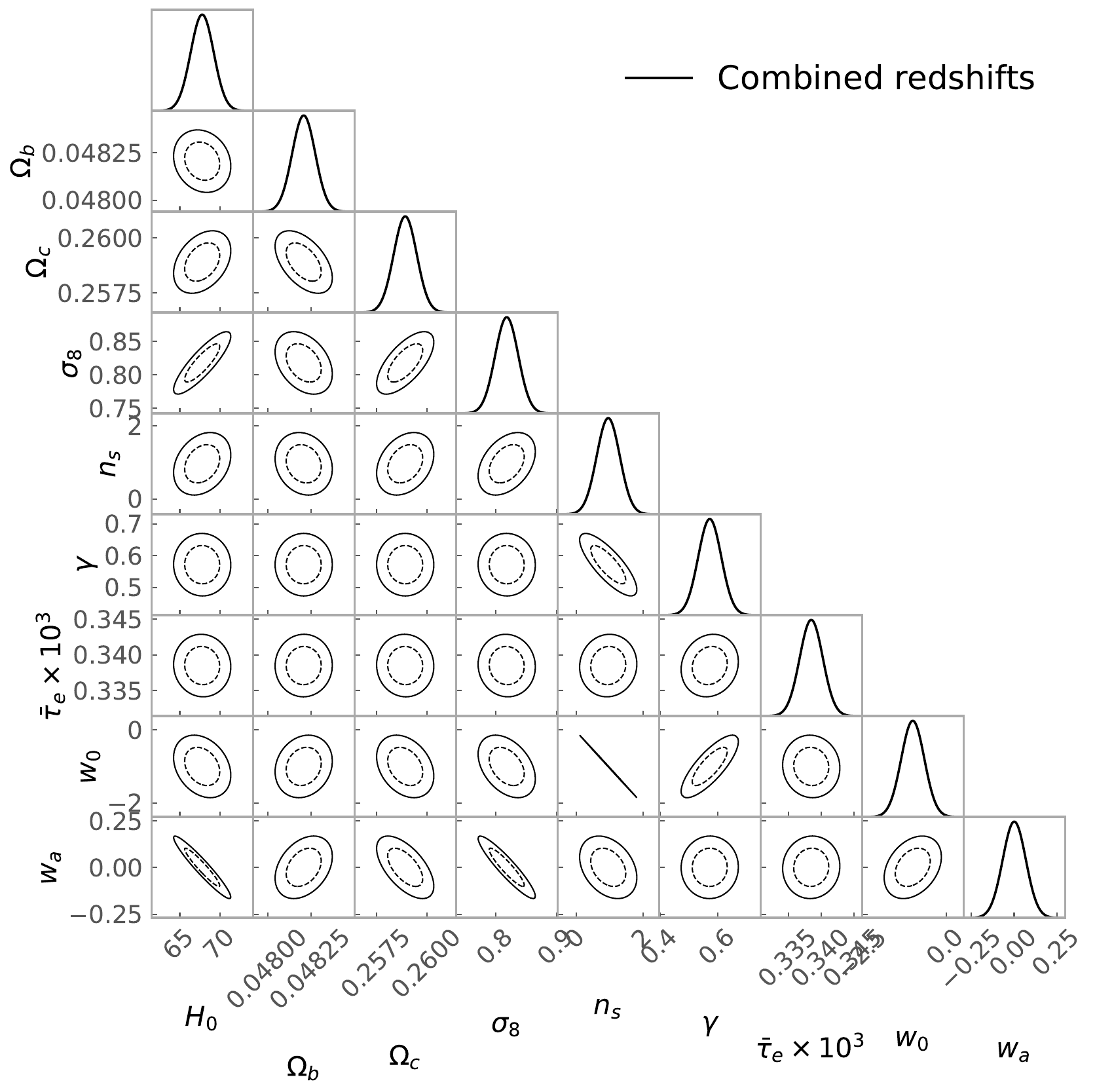} \label{fig:cosmo_all}} 
    \caption{Constraints for all the cosmological parameters considered in this work. The cosmological model \wcdm + $\gamma$ + $w_a$ is divided into: (a) the redshift dependent part of the constraints, where there are very relaxed priors to show the constraining potential of pairwise kSZ alone on cosmology; and (b) using \Planck-TTTEEE-low$\ell$-lowE-BAO \wcdm+$w_a$ 2018-like priors for all the \lcdm parameters and the Compton-y priors for $\bar{\tau}_e$.}
\end{sidewaysfigure*}

Here we show the effects of all the cosmological parameters that we considered for the pairwise kSZ analysis. This is a cosmological model of \lcdm + $\gamma$ + $w_0$ + $w_a$, where we want to observe also the redshift dependent constraints that can be achieved by using a redshift bin separation of 0.2. The redshift dependence is shown in Fig.~\ref{fig:cosmo_zbins_all}, where we did not include priors from \Planck on the parameters, allowing us to see the constraining power of purely the pairwise kSZ at a binned redshift level. The full combined constraints are presented in Fig.~\ref{fig:cosmo_all}, where \Planck's covariance from the low$\ell$+low$E$+BAO were used as priors for the main \wcdm+$w_a$ parameters, and the uncertainty on $\bar{\tau}_e$ that comes from the y-map was used as a prior for this parameter. We show the constrained parameters in Tab.~\ref{tab:cosmo_all}, where we observe how most parameters are not changed at all by adding the pairwise kSZ, but $\gamma$ can still be constrained at the $14\sigma$ level.

\begin{table*}
    \centering
    \begin{tabular}{ccccc}
        \hline
        \hline
        Parameter & Fiducial & $\sigma_{\rm \Planck}$ & $\sigma_{\Planck + pkSZ}$ & Improvement over \Planck\\
        \hline
        $H_0$ & 67.77 & 1.44 & 1.42 & 1\% \\
        $w_{b 0}$ & 0.048 & $6.65\times 10^{-5}$ & $6.64 \times 10^{-5}$ & $<1\%$ \\
        $w_{dm 0}$ & 0.259 & $5.74 \times 10^{-4}$ & $5.72 \times 10^{-4}$ & $<1\%$ \\
        $\sigma_8$ & 0.818 & $1.91 \times 10^{-2}$ & $1.88 \times 10^{-2}$ & 2\% \\
        $n_s$ & 0.96 & $3.45 \times 10^{-1}$ & $3.45 \times 10^{-1}$ & - \\
        $\gamma$ & 0.57 & 0.57 & 0.04 & 93\% \\
        $w_0$ & -1 & 0.34 & 0.34 & - \\
        $w_a$ & 0 & 0.06 & 0.06 & - \\
        \hline
        \hline
    \end{tabular}
    \caption{\lcdm + $\gamma$ + $w_0$ + $w_a$ cosmological parameters constraints after combining the redshift bins. The fiducial values come from \Planck-TTTEEE-low$\ell$-lowE-BAO \wcdm+$w_a$ 2018 unceratinties. The constrained value is the one obtained from the CMB-S4 pairwise kSZ uncertainties. The improvement represents the percentage change when adding the pairwise kSZ measurement with respect to the \Planck cosmological constraints.}
    \label{tab:cosmo_all}
\end{table*}

\end{document}